# Dominating Interlayer Resonant Energy Transfer in Type-II 2D Heterostructure

*Arka Karmakar[1]\*, Abdullah Al-Mahboob[1], Christopher E. Petoukhoff[1], Oksana Kravchyna[1], Nicholas S. Chan[1], Takashi Taniguchi[2], Kenji Watanabe[3], Keshav M. Dani[1]*

[1] Femtosecond Spectroscopy Unit, Okinawa Institute of Science and Technology Graduate University, 1919-1 Tancha, Onna, Kunigami District, Okinawa 904-0495, Japan

[2] International Center for Materials Nanoarchitectonics, National Institute for Materials Science, 1-1 Namiki, Tsukuba, Ibaraki 305-0044, Japan

[3] Research Center for Functional Materials, National Institute for Materials Science, 1-1 Namiki, Tsukuba, Ibaraki 305-0044, Japan

\*Corresponding Author: arka.karmakar@oist.jp; karmakararka@gmail.com

**ABSTRACT**

Type-II heterostructures (HSs) are essential components of modern electronic and optoelectronic devices. Earlier studies have found that in type-II transition metal dichalcogenide (TMD) HSs, the dominating carrier relaxation pathway is the interlayer charge transfer (CT) mechanism. Here, this report shows that, in a type-II HS formed between monolayers of $MoSe_2$ and $ReS_2$, nonradiative energy transfer (ET) from higher to lower work function material ($ReS_2$ to $MoSe_2$) dominates over the traditional CT process with and *without* a charge-blocking interlayer. Without a charge-blocking interlayer, the HS area shows 3.6 times $MoSe_2$ photoluminescence (PL) enhancement as compared to the MoSe$_2$ area alone. After completely blocking the CT process, more than one order of magnitude higher MoSe$_2$ PL emission was achieved from the HS area. This work reveals that the nature of this ET is truly a resonant effect by showing that in a similar type-II HS formed by ReS$_2$ and WSe$_2$, CT dominates over ET, resulting in a severely quenched WSe$_2$ PL. This study not only provides significant insight into the competing interlayer processes, but also shows an innovative way to increase the PL quantum yield of the desired TMD material using ET process by carefully choosing the right material combination for HS.

**INTRODUCTION**

Energy transfer (ET) is a process in which energy is nonradiatively transferred from an excited fluorophore (donor) to another fluorophore (acceptor).[1–3] ET is a long-range process without involving emission and reabsorption of photons.[4] The ET theory is based on the concept considering an excited fluorophore as an oscillating dipole transferring energy to another dipole; similar as coupled pendulums in classical mechanics. We can use the following equation to describe the ET process from donor (D) to acceptor (A):

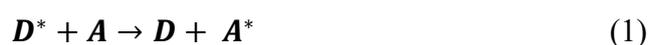

$$D^* + A \rightarrow D + A^* \qquad (1)$$

where $D^*$ and $A^*$ are the excited states of the fluorophores. When donor and acceptor are in close proximity and donor emission spectra overlaps with the acceptor absorption spectra, the excited state energy from the donor gets transferred to the acceptor via dipole-dipole interaction, and subsequently released in the form of radiative energy emission from the acceptor.[2,3] ET, thus favorably occurs between parallel dipoles.[3] ET has already been proven as building block of many biological and chemical applications.[5–8] Despite the long history of discovery,[1] ET exploration is still on-going due to the fact that many promised applications are based on the idea of placing donor and acceptor in sub-nanometer proximity, to fully utilize the ET process.[9]

Layered transition metal dichalcogenide (TMD) materials can be exfoliated down to atomically-thin monolayers (1Ls) and the optical bandgaps of these materials spans the broad range of near-infrared to deep ultraviolet in the electromagnetic spectrum.[10] Sub-nanometer spacing in heterostructures (HSs) created by stacking different van der Waals (vdW) materials has enabled new opportunities to study exciting phenomena in 2D systems, such as Moiré patterns, valleytronics, spintronics and interlayer excitons just to mention a few.[11–14] Thus, TMD HSs are ideal candidates to study the ET process because they can be stacked in atomically close proximity. In TMD HSs, interlayer energy and charge transfer (CT) compete with each other; thus, comprehensive understanding of these processes is necessary to develop TMD-based applications.

The majority of the available TMD materials form type-II HSs.[15] In traditional type-II TMD HSs, the dominating charge carrier relaxation pathway is interlayer CT process[16–18] rather than interlayer ET mechanism. Kozawa *et al.*[19] showed the presence of interlayer ET in type-II HS by placing an atomically thin charge-blocking layer, namely a few-layer hexagonal boron nitride (hBN) in between the two TMD layers to suppress the interlayer CT process. Studying efficient ET process in TMD HSs requires materials with coinciding donor emission and acceptor absorption peak positions. Selection of the right TMD pairs is thus crucial to observing ET in type-II HSs.

A recent study[20] has predicted that a type-II HS formed between 1L rhenium disulfide (ReS$_2$) and 1L molybdenum diselenide (MoSe$_2$) integrated on a flexible substrate is an efficient candidate for near infrared (NIR) photodetection. Here, for the first time we experimentally study this unique HS and show that interlayer ET from ReS$_2$ to MoSe$_2$ dominates over the traditional interlayer CT process. As the donor, the emission of ReS$_2$ overlaps with the absorption of the acceptor, MoSe$_2$, and ReS$_2$ has dominating nonradiative recombination channel. We show that the nonradiative ET from ReS$_2$ *resonantly* excites more carriers in MoSe$_2$ giving a 3.6 times enhanced PL emission from the HS even *without* having a charge-blocking layer in between the two materials. By adding different thicknesses of hBN interlayer, we found that the PL enhancement, which is proportional to the ET, has a distance (d) dependence of ~1/d$^2$ at larger separation, indicating 2D dipole-2D dipole interaction between the TMD layers.[21–23] At donor-acceptor separation of 13.5 nm we successfully achieved 12.5 times higher MoSe$_2$ PL emission from the HS area by completely suppressing the other competing processes. We further show that, by replacing the MoSe$_2$ layer with 1L tungsten diselenide (WSe$_2$), the resonant nature of ET breaks and interlayer CT dominates in the WSe$_2$-ReS$_2$ HS like other typical type-II HSs, exhibiting a severely quenched WSe$_2$ PL emission from the HS. This report shows that by carefully choosing the right material combination, we can interplay between the ET and CT processes, and thus enhance the PL efficiency of the desired TMD material using resonant ET.

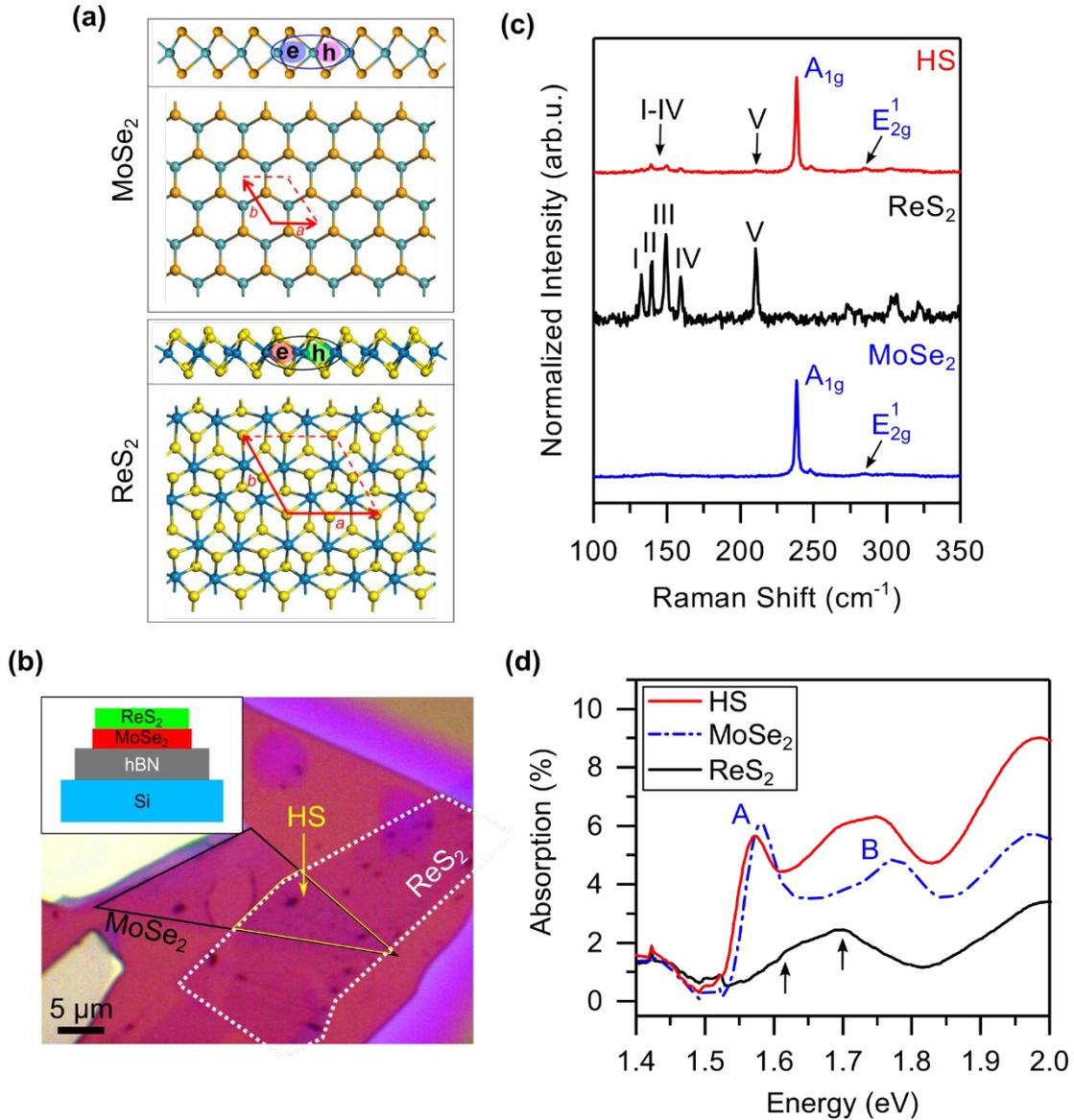

**Figure 1.** (a) Monolayer (1L) MoSe$_2$ and ReS$_2$ crystal structure. Top panels show the side view and bottom panels show the top view of the crystal structures. Side views show the schematic of in-plane orientation of dipoles across these layered materials. (b) Optical image of the ReS$_2$-MoSe$_2$ heterostructure from Sample 1 (S1). Inset is the schematic illustration of the sample's side view. (c) Raman spectra from the MoSe$_2$, ReS$_2$ and HS area. HS Raman spectra consists of different vibrational modes from individual 1L area. (d) Absorption spectroscopy data of three different areas from a similar heterostructure made on a transparent sapphire substrate (Sample 2, S2). MoSe$_2$ A and B excitonic peaks are clearly visible and ReS$_2$ lower energy absorption peaks are marked with arrows. HS spectra consists of peaks from both 1L areas.

**RESULTS AND DISCUSSION**

The atomic arrangement of in-plane crystal structures of 1L MoSe$_2$ and 1L ReS$_2$ are shown in Figure 1(a), where the top and bottom panels of MoSe$_2$ (or ReS$_2$) respectively show the side and top views of the materials. The side views also show the schematic illustration of the orientation of in-plane dipoles in these layered materials. MoSe$_2$ crystal structure has perfect in-plane 3-fold symmetry, whereas, ReS$_2$ is an anisotropic material having triclinic crystal structure with oblique in-plane lattice,[24,25] as shown in Figure 1(a). 1L ReS$_2$-1L MoSe$_2$ HS samples were fabricated using micromechanical exfoliation (see *Experimental Details*) and were stacked onto thick hBN layers to eliminate any effect from surface-mediated trap states during our experiment. Figure 1(b) shows the optical micrograph of the fabricated 1L ReS$_2$-1L MoSe$_2$ HS on hBN/Si substrate, namely Sample 1 (hereafter S1), where the inset shows a schematic illustration of the cross-section (stacking of layers) of the fabricated sample. The representative Raman spectra (Figure 1(c)) from the three different areas on S1 show MoSe$_2$ characteristics A$_{1g}$ and E$^1_{2g}$ peaks,[26–28] ReS$_2$ intralayer modes (I-V),[29,30] and the HS Raman spectra consists of the characteristic peaks from both materials. It is also worth to note that, in the HS Raman spectra, the signature peak positions of each material remain the same as compared to the individual monolayer area, suggesting a similar level of doping and strain in the HS area as compared to the monolayer areas.[31,32] Figure 1(d) shows the absorption spectra from another HS made on a transparent sapphire substrate, Sample 2 (S2). We observed the characteristic A and B excitonic absorption from MoSe$_2$,[19] as well as two identifying absorption peaks in ReS$_2$ within close proximity[33] (pointed with arrows in Figure 1(d)). HS absorption spectrum consists of peaks from each layer and showed an overall higher optical absorption as compared to the monolayer areas. The slight redshift in the HS absorption peaks as compared to the MoSe$_2$ area could be due to the change in electronic coupling and dielectric environment.[34,35]

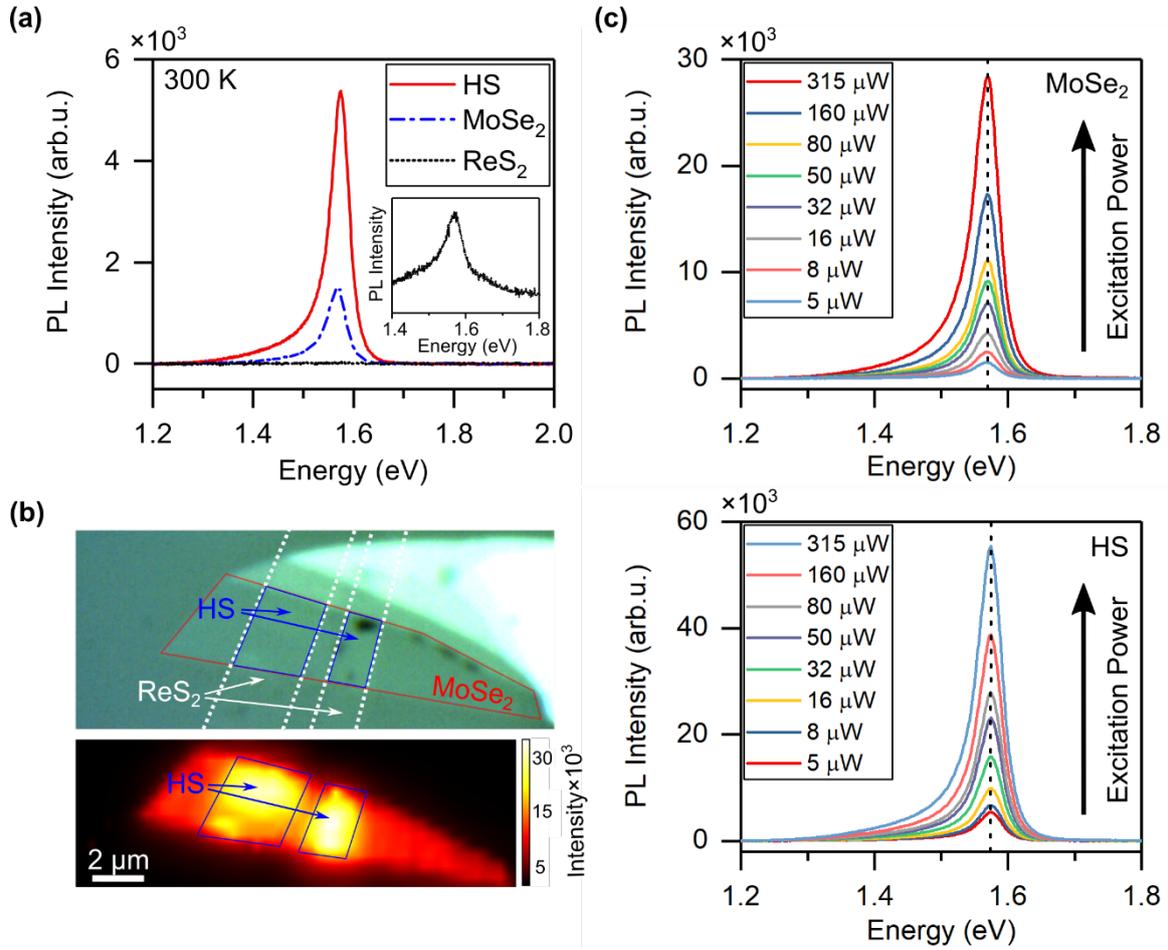

**Figure 2.** (a) Room temperature (RT) PL emission from the MoSe$_2$, ReS$_2$ and HS area of S1. Inset is the ReS$_2$ PL emission obtained at ~30 times higher laser power and longer accumulation. (b) Optical micrograph of another ReS$_2$-MoSe$_2$ HS, Sample 3 (S3) and PL intensity map from the same sample. The entire HS area shows this enhanced PL emission. (c) Laser power dependent PL from the MoSe$_2$ and HS area of S1. PL peak energy does not shift with increasing laser power, indicating that throughout the experiment we remain in material's linear regime.

Room temperature (RT) PL peak position obtained from the MoSe$_2$ area of S1 matches well with previous published results[26,27,36] (Figure 2(a)). Being a pseudo-indirect bandgap semiconductor and having extremely low PL quantum yields, 1L ReS$_2$ has almost non-existent PL emission[33,37,38] (Figure 2(a)). To obtain analyzable signal-to-noise ratio in the ReS$_2$ PL spectrum, the laser excitation power required was ~30 times higher, and the accumulation time was much longer compared to the MoSe$_2$ and HS areas (inset of Figure 2(a)). It is important to observe that at RT, ReS$_2$ PL emission and MoSe$_2$ A excitonic absorption peak position almost *coincided* with each other around 1.60 eV. Such coincidence between the nonradiative and radiative excitonic bandgaps enables resonant ET to dominate, as discussed in later sections. Although 1L ReS$_2$ had slightly lower absorption than 1L MoSe$_2$ (Figure 1(d)), the almost nonexistent PL emission further proved that nonradiative recombination dominated in the 1L ReS$_2$ film. The Raman peak positions and strong (weak) PL intensity of MoSe$_2$ (ReS$_2$) prove their monolayer nature. The HS PL peak position perfectly matched with the MoSe$_2$ PL peak position, but the luminescence from the HS area was enhanced by a factor of ~3.6 as compared to the PL emission from the area of MoSe$_2$ alone (Figure 2(a)). The enhancement factor considered here is the ratio of the HS PL peak intensity to the MoSe$_2$ peak intensity, when the two area are excited by the same low illumination intensity and accumulation time. In order to check the distribution of this enhanced PL emission over the HS area, we collected a PL intensity map on a similar ReS$_2$-MoSe$_2$ HS, namely Sample 3 (hereafter S3) (Figure 2(b)) (see S.I. Figure S1 for the PL spectra of the sample). The red area in Figure 2(b) corresponds to the 1L MoSe$_2$ area. From the PL map, it is clear that the PL enhancement was not a localized phenomenon, but the entire HS area showed this enhanced PL emission at RT. To further investigate the origin of this enhanced PL emission from the HS (S1), we performed low temperature (LT) PL measurement at 100 K (Figure S2). At LT, the HS still showed a similar PL enhancement factor of ~3, indicating that weakly temperature dependent dipole-dipole coupling dominated in this HS rather than the Dexter-type ET process (interlayer exciton transfer).[19,39] Dexter-type ET can also be described using Equation 1, but the mechanism is different as discussed later. There is another possibility of the contribution of constructive interference from the back-reflected light at the sample-substrate interface as the origin of this enhanced PL emission.[40] However, we observed the HS PL enhancement in all of our samples irrespective of the different hBN substrate thicknesses, suggesting this interference was not a contributing factor. To further confirm this, we performed optical transfer matrix method (TMM) calculations at different hBN thicknesses, and the results ruled out the possibility of any notable contribution from the optical interference effect in comparison to the observed PL enhancement (see S.I. for detailed discussion and Figure S3). Excitation power dependent PL measurements confirmed that we remained within the linear regimes of the materials (Figure 2(c)): for both the MoSe$_2$ and HS areas, the peak energy and full width at half maxima (FWHM) of the PL emission (Figure S4) remained unchanged with increasing excitation power. To check the effect of dielectric screening as the origin of this PL enhancement[41] in the encapsulated MoSe$_2$ layer, we fabricated an inverted MoSe$_2$-ReS$_2$ HS and observed similar PL enhancement (see Figure S5 for optical image and PL spectra of the sample), proving that dielectric screening had negligible effects in this study.

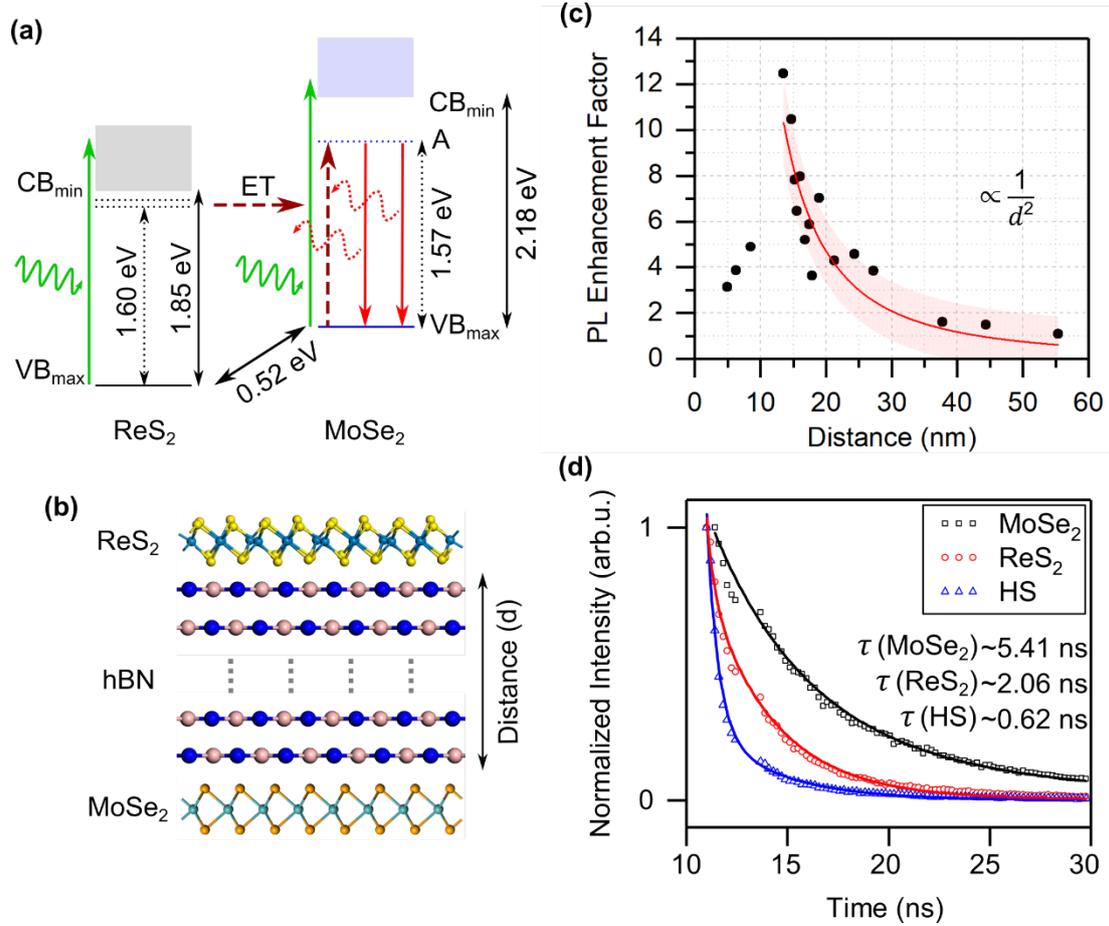

**Figure 3.** (a) Energy diagram representing the PL enhancement mechanism due to the resonance excitonic energy transfer from the ReS$_2$ to MoSe$_2$ layer. Excitonic levels are shown as dotted line below the conduction band minima. (b) Schematic illustration of the HS samples made with different thicknesses (distance) of hBN interlayer. (c) Change of PL enhancement from the HS area with distance. Pl enhancement factor shows $1/d^2$ dependency at larger separation. (d) RT time resolved PL (TR-PL) spectra from the MoSe$_2$, ReS$_2$ and HS area of D3. Curve fitted data are shown by the solid line.

In order to experimentally determine the HS band-alignment type, we used kelvin probe force microscopy (KPFM) to obtain the valence band (VB) offset ($\Delta E_V$) between the MoSe$_2$ and ReS$_2$ layers as ~0.52 eV (Figure S6 and S7). Combining the experimental value of $\Delta E_V$ obtained from the KPFM measurement with the electronic bandgap,[42,43] we determined that 1L ReS$_2$-1L MoSe$_2$ HS indeed form a type-II band alignment (Figure 3(a)) in agreement with earlier published theoretical works.[15,20] This eliminated the possibility of short-range Dexter-type ET in our system, as it requires the complete exciton to be transferred from donor to acceptor, which can only be accomplished in type-I HSs. In typical type-II HSs, the PL emission of both materials quench after forming the HS due to the interlayer CT process.[16,17,19,44] Interestingly the ReS$_2$-MoSe$_2$ HS, exhibited *enhanced* MoSe$_2$ PL emission from the HS area. We conclude that this enhancement was due to the ET process from the ReS$_2$ layer to the MoSe$_2$ layer. One possible reason of this unusual dominating ET process could be the suppression of coherent CT processes in this HS due to the weaker electronic coupling associated with symmetry mismatch and huge lattice mismatch between the trigonal in-plane of 2*H*-MoSe$_2$ and oblique in-plane of 1*T*-ReS$_2$ structure in comparison with the most studied 2*H*-TMD HSs. A recent ultrafast study on MoS$_2$-ReS$_2$ HS,[45] showed much slower photocarrier transfer time on the order of 1 ps, as compared to the ~100 fs CT time in the HS formed between 2*H*-2*H* crystals.[17,46,47] However, the quantifiable reason of this dominating ET process remains unclear. Figure 3(a) shows the schematic illustration of the PL enhancement due to the proposed ET mechanism in the ReS$_2$-MoSe$_2$ HS. Upon photoexcitation with 2.33 eV (532 nm), electrons in the TMD materials are excited from VB to CB, leaving holes in the VB. The excited electrons quickly relax to the excitonic states situated slightly lower than the CB minimum[43,48] and subsequently recombine at the VB releasing photons of the energy equal to the optical bandgap. In ReS$_2$, the majority of excited state carriers nonradiatively recombine to the VB via phonon-mediated processes. We propose the following mechanism to explain the PL enhancement: in the presence of weaker interlayer CT process, more carriers in MoSe$_2$ layer resonantly excite in the excitonic level due to instantaneous ET from the ReS$_2$ layer, giving a significant excess of radiative excitons in MoSe$_2$. Thus, the MoSe$_2$ PL emission is enhanced in the HS as shown in Figure 3(a). We excluded the possibility of the ReS$_2$ PL enhancement in the HS due to the fact that the HS PL emission shape and peak position perfectly matched with the MoSe$_2$ PL spectra (Figure S8). To understand the nature of the ET process, we made several HSs with varying thickness of hBN layers inserted between the TMD layers (Figure 3(b) and S9). ET shows a distance dependence between the donor and acceptor as $1/d^n$, where n is a factor dependent on the systems dimensionality:[49] n = 6 for 0D-to-0D dipoles coupling, n = 4 for 0D-to-2D dipoles coupling and n = 2 for 2D-to-2D dipoles coupling.[21,22] It is important to mention that, at shorter distances several processes such as charge tunneling through the hBN barrier, interlayer ET process and local impurity related opacity at the hBN-TMD interface could compete with each other. Thus, we considered to observe the pure ET effect from the maximum enhancement point in our experiment (Figure 3(c)). Specifically, at a distance of 13.5 and 14.7 nm we observed more than one order of magnitude higher PL enhancement factor of ~12.5 and 10.5, respectively. Beyond the maximum enhancement point, PL enhancement factor, which is considered to be directly proportional to ET, decreased with a

~1/d² dependency in our HS systems, which suggests that ET from $ReS_2$ to $MoSe_2$ occurred via 2D-to-2D dipoles interaction.[21–23] The fluctuation observed in the data within 15-25 nm distance (Figure 3(c)), could the combined effect of varying material's property from sample-to-sample and the change in coupling due to the rotational mismatch in our randomly stacked samples. A detailed understanding of the twist angle dependence in the ET process requires further study. We further confirmed the long-range dipole-dipole ET mechanism through a shortening of the excited state lifetime of the HS compared to the individual layers (Figure 3(d)). Compared to the isolated monolayers, which had PL lifetimes of ~ 5.41 and 2.06 ns for the acceptor ($MoSe_2$) and donor ($ReS_2$), respectively, the donor lifetime in the HS was shortened to ~620 ps (see *Experimental Details*). ET occurs through a decreased donor fluorescence intensity and reduction of excited state lifetime accompanied by an increased acceptor fluorescence intensity. We find that there was a significant reduction of $ReS_2$ PL decay time in the HS as compared to the $ReS_2$ area alone (Figure 3(d)), which further proves that energy was transferred from $ReS_2$ to $MoSe_2$. To find the ET efficiency ($E_{ET}$) and ET rate ($k_{ET}$) in the $ReS_2$-$MoSe_2$ HS, we used the following equations:[4,50,51]

$$E_{ET} = 1 - \frac{\tau_{ReS2,HS}}{\tau_{ReS2}} \quad (2)$$

$$k_{ET} = \frac{1}{\tau_{ReS2,HS}} - \frac{1}{\tau_{ReS2}} \quad (3)$$

where $\tau_{ReS2, HS}$ is the $ReS_2$ PL lifetime in HS. By using the above-mentioned values in Equation 2 and 3 we calculate the ET efficiency and rate in the $ReS_2$-$MoSe_2$ HS to be ~70% and ~1.13 ns$^{-1}$ respectively. It is important to note that the traditional ET understanding is based on the model which considers the coupling between optically emissive (bright) dipoles only.[52,53] A recent study[54] has shown that, ET from optically non-emissive (dark) excitons dominate in TMD HSs. Considering the extremely low PL quantum yield in $ReS_2$ (~10$^{-4}$),[38] dark excitons dominate. Thus, we speculate that the majority of ET from $ReS_2$ dark excitons resonantly enhanced the radiative exciton population in $MoSe_2$ giving 360% PL enhancement despite the presence of interlayer CT process. Measuring the ET efficiency of the dark excitons requires further investigation in future work.

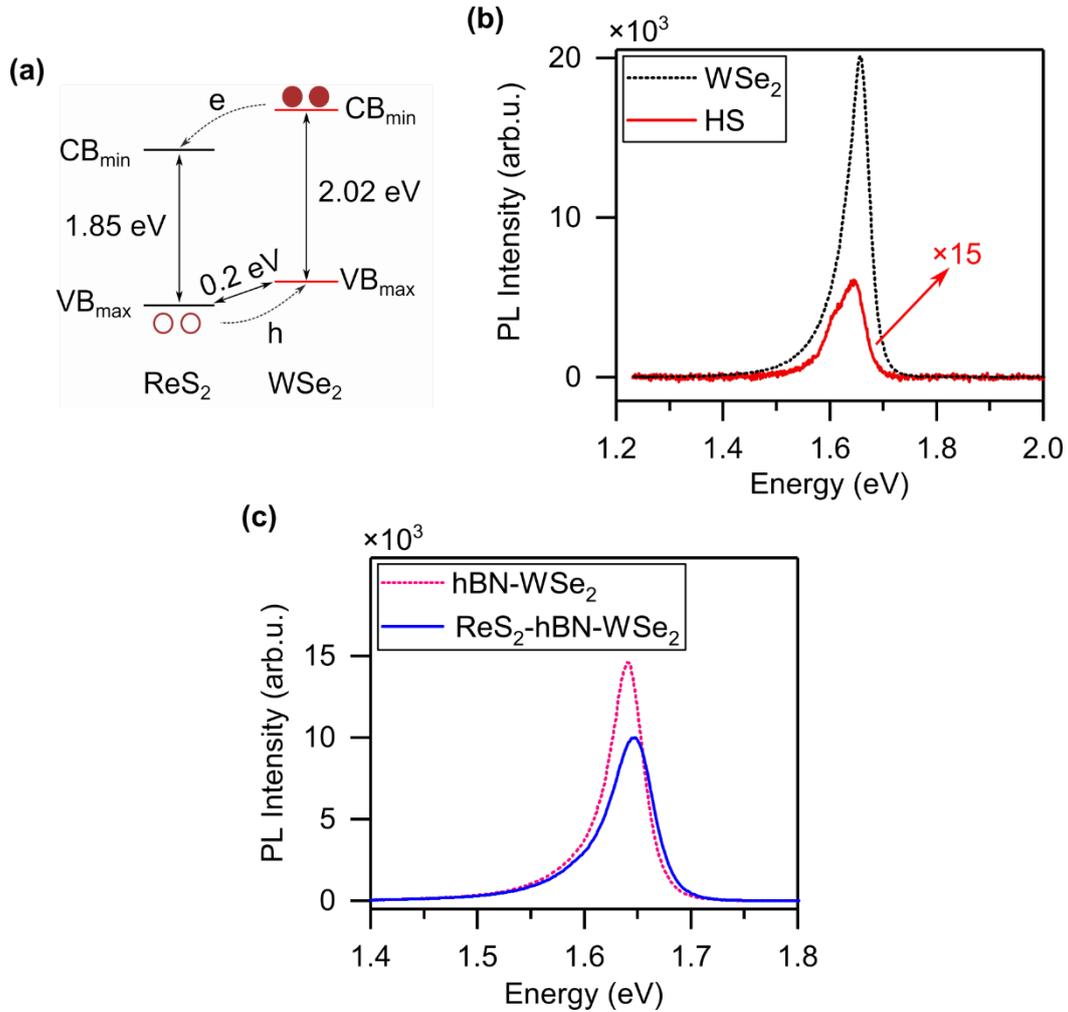

**Figure 4.** (a) Schematic illustration of type-II band alignment of ReS$_2$-WSe$_2$ HS. (b) PL measurement taken from the ReS$_2$-WSe$_2$ HS (Sample 4, S4) and 1L WSe$_2$ area. HS PL spectra shows massive quenching of WSe$_2$ emission. (c) PL spectra of the ReS$_2$-hBN-WSe$_2$ HS (Sample 5, S5) before and after transferring the top ReS$_2$ layer. After transferring the top ReS$_2$ layer, HS does not show an enhanced PL emission.

In order to check the resonant nature of the interlayer ET process, we replaced the MoSe$_2$ layer with WSe$_2$ (Sample 4, S4), which has about 80 meV larger optical bandgap than MoSe$_2$ (Figure 4 and S.I. Figure S10). The experimentally obtained $\Delta E_V$ between ReS$_2$ and WSe$_2$ from the KPFM measurement to be ~0.2 eV (Figure S11). Combining the $\Delta E_V$ value with WSe$_2$ electronic bandgap,[55] we found that 1L ReS$_2$-1L WSe$_2$ also form a similar type-II HS (Figure 4(a)), in agreement with previous reports.[15,20] PL measurements showed severely quenched HS PL emission (Figure 4(b)) due to interlayer CT being the dominant mechanism (Figure 4(a)), which is typical in type-II TMD HSs. It is worth mentioning here that in these ReS$_2$-MoSe$_2$ and ReS$_2$-WSe$_2$ HSs, the traditional type-II interlayer PL emissions are not visible due to the valley-forbidden dark excitonic nature (ReS$_2$ excitonic transition happens around the Γ valley in contrast with K valley in 2$H$ materials).[20] To eliminate the interlayer CT process, we prepared a ReS$_2$-WSe$_2$ HS (Sample 5, S5) with few layers of hBN inserted in between. PL spectra of this HS did not lead to enhanced emission after placing the top ReS$_2$ layer (Figure 4(c)). The slightly reduced PL emission from the WSe$_2$ layer after placing the ReS$_2$ was mainly attributed to the absorption of the incoming photons from the top layer and scattering at the multiple interfaces. This was in stark contrast to the ReS$_2$-MoSe$_2$ and ReS$_2$-hBN-MoSe$_2$ samples, where we observed PL enhancement regardless of the presence of a charge blocking hBN interlayer. It is evident that the ET from the ReS$_2$ layer was insufficient to enhance the PL emission from WSe$_2$ layer and the interlayer CT process dominated in this HS.

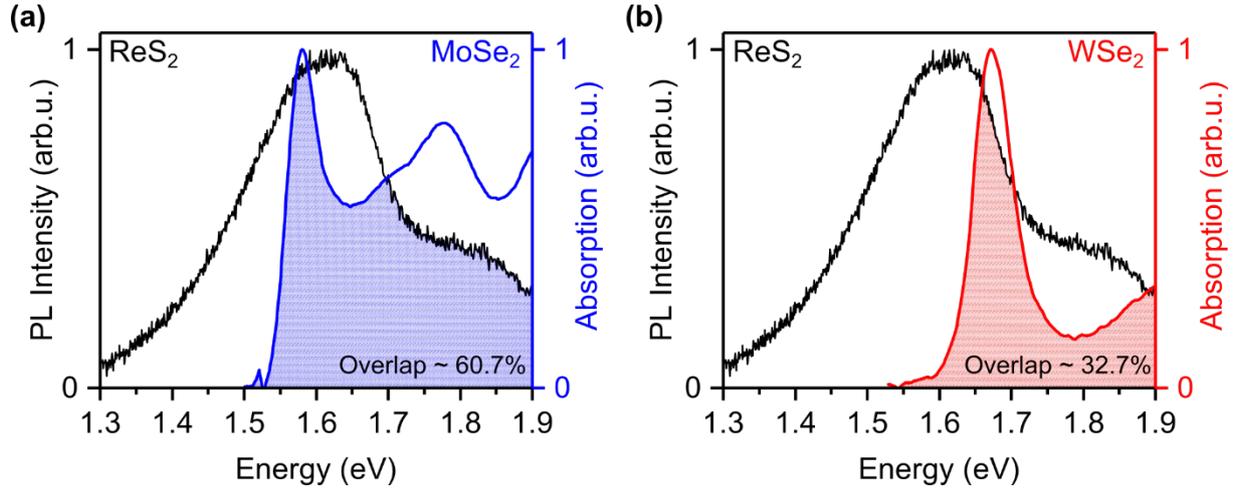

**Figure 5.** (a) and (b) Area overlaps between the normalized ReS$_2$ (donor) PL emission and the MoSe$_2$, WSe$_2$ (acceptor) absorption spectra at RT, respectively. The area overlaps for the MoSe$_2$ and WSe$_2$ system are ~60.7% and ~32.7%, respectively.

To further investigate the nature of the ET in TMDs, we examine the spectral overlap of the emission and absorption of donor and acceptors (Figure 5). The spectral area overlaps with respect to ReS$_2$ emission for MoSe$_2$ and WSe$_2$ were ~60.7% and ~32.7%, respectively. It has been previously shown experimentally that $E_{ET}$ is proportional to the spectral overlap between the donor emission and acceptor absorption spectra.[56] Thus, based on the spectral overlaps, ET should be less efficient in the ReS$_2$-WSe$_2$ HS. However, the massive quenching of the WSe$_2$ PL emission from the HS proves that the ET rate cannot compete with the CT rate and donor relaxation rate in this configuration. This finally proves that nonradiative excitonic ET from ReS$_2$ *resonantly* excites radiative excitons in MoSe$_2$, resulting in the significant PL enhancement.

**CONCLUSION**

In summary, we have shown, to the best of our knowledge, for the first-time interlayer resonant ET process in a type-II TMD HS *without* a charge blocking interlayer, resulting in 3.6 times MoSe$_2$ PL enhancement from the HS. We calculated the ET efficiency from the bright excitons in our ReS$_2$-MoSe$_2$ HS to be ~70%, with a potential for increasing the overall efficiency in future work. The experimental findings suggest that both the dark and bright excitonic ET responsible for the enhanced PL emission and the resonant ET contribute dominantly to form bright excitons in MoSe$_2$ and promote radiative recombination in MoSe$_2$ at the HS region. Through insertion of a charge blocking hBN layer, we effectively suppress the competing CT process, achieving a maximum PL enhancement factor of 12.5 at a distance of 13.5 nm between the TMD layers. By varying the separation distance between layers, we demonstrated the long-range ET has a distance dependence of ~$1/d^2$ for larger separation, indicating 2D dipoles to 2D dipoles coupling between ReS$_2$ and MoSe$_2$ layers. Finally, we showed that by replacing the MoSe$_2$ layer with WSe$_2$, the resonant nature of ET broke, leading to interlayer CT dominating over the ET process, which resulted in a severely quenched WSe$_2$ PL emission from the HS. The ReS$_2$-MoSe$_2$ HS provides a unique opportunity to simultaneously study the ET and CT processes in TMD materials at ultrafast timescale in future work. We strongly believe that this work will pave the way to discover more HS combinations that can exploit the normally parasitic nature of non-emissive 2D materials by increasing the PL quantum yield of 2D semiconducting emitters via ET, which can lead to improved TMD-based photovoltaic, optoelectronic, and photonic devices.

**EXPERIMENTAL DETAILS**

**HS fabrication:** hBN and TMD crystals were obtained from the National Institute of Materials Science, Japan; HQ Graphene, Netherlands and 2D Semiconductors, USA, respectively. hBN was exfoliated using the standard mechanical exfoliation technique. 1L TMD materials were exfoliated using the standard PDMS-based technique and carefully stacked on top of hBN flake using a homebuilt semi-automated transfer stage. $ReS_2$-$MoSe_2$ and $ReS_2$-$WSe_2$ HSs were annealed under high vacuum (~5×$10^{-6}$ Torr) at 250 ºC for 4-6 hours to remove any polymer residue and make the monolayers in good contact with each other.

**Characterization:** For the **AFM** measurement, we used Bruker Dimension ICON3-OS1707 in 'ScanAsyst' (peak force tapping) mode to obtain the highest resolution AFM image. We performed the **PL** and **Raman** spectroscopy measurement using Tokyo Instruments' Nanofinder 30 Microspectrometer. For low-temperature PL measurement the sample was loaded and cooled in a cryostat with continuous flow of liquid-nitrogen ($N_2$). For PL measurement the sample was illuminated using CW solid-state 532 nm laser with an average power of 5 µW (spot size ~1 µm) and focused by a 100x objective lens (N.A. 0.95). For the **absorption** spectroscopy analysis, we used Craic 20/30 PV Microspectro-photometer system with white light illumination at perpendicular direction on the sample surface. The transmitted light passing through a variable aperture were collected using a 100x objective lens (N.A. 0.9) before entering to the spectrum analyzer. The transmission spectra was converted to absorption spectra using equation, $A\% = 100 - T\%$ (considering no reflection in the system).

We performed the **KPFM** measurement using the Bruker MultiMode 8 system with a platinum-iridium coated conductive tip. To obtain the high-resolution data we performed the measurement in frequency modulation (FM-KPFM) mode, which applies an AC voltage to the probe at very low frequency to obtain the electrostatic potential map of the sample surface.

**TR-PL** measurement was performed using a Spectra-Physics Matai XF-IMW femtosecond Ti:sapphire laser with tunable wavelengths from 710-920 nm, pulse width of 70 fs, and repetition rate of 80 MHz the 800 nm fundamental beam was passed through a second harmonic generator and pulse picker module to generate an excitation source with 400 nm wavelength and 4 MHz repetition rate. The spot size of the excitation beam on the sample surface was approximately 3 µm, with an average power of 90 µW. PL from the sample was collected through another 100x microscope objective (N.A. 0.70) and directed into a streak scope (Hamamatsu Photonics, C10627-03) coupled to a CCD. The instrument response function was determined by measuring the scattered laser light, which gave a temporal resolution of ~15 ps. After subtracting the instrumental response and the rise time, we fitted the normalized data from the $MoSe_2$ and $ReS_2$ monolayers area with a single-exponential decay function to get the faster decay component (intrinsic PL lifetime) . To obtain the $ReS_2$ lifetime in the HS, we fitted the HS data using the following equation:[57]

$$f(x) = A_1 [exp(-(x-x_0)/\tau_{ReS2,HS}) - exp(-(x-x_0)/\tau_{MoSe2,HS})] + A_2 [exp(-(x-x_o)/\tau_{MoSe2})]$$

where $A_1$, $A_2$ are constants, $\tau_{ReS2,HS}$ is the $ReS_2$ PL lifetime in the HS, $\tau_{MoSe2,HS}$ is the PL lifetime of excess charge in the HS $MoSe_2$ due to ET and $\tau_{MoSe2}$ is the $MoSe_2$ PL lifetime.

**Acknowledgements**

AK acknowledges the useful discussion with Chakradhar Sahoo. This work was supported by the funding from the Femtosecond Spectroscopy Unit at the Okinawa Institute of Science and Technology Graduate University. K.W. and T.T. acknowledge support from the Elemental Strategy Initiative conducted by the MEXT, Japan (Grant Number JPMXP0112101001) and JSPS KAKENHI (Grant Numbers 19H05790 and JP20H00354).

**Conflict of Interest**

The authors declare no conflict of interest.

**Supporting Information**

RT and LT PL spectra, details of TMM calculation, power dependent PL spectroscopy analysis, KPFM potential map and AFM height profile, Figure S1-S11.

**REFERENCES**

(1) Förster, T. Energy Migration and Fluorescence. *J. Biomed. Opt.* **2012**, *17* (1), 011002.

(2) Koushik, S. V; Vogel, S. S. Energy Migration Alters the Fluorescence Lifetime of Cerulean: Implications for Fluorescence Lifetime Imaging Forster Resonance Energy Transfer Measurements. *J. Biomed. Opt.* **2008**, *13* (3), 1–9. https://doi.org/10.1117/1.2940367.

(3) Clegg, R. M. Fluorescence Resonance Energy Transfer. *Curr. Opin. Biotechnol.* **1995**, *6* (1), 103–110. https://doi.org/https://doi.org/10.1016/0958-1669(95)80016-6.

(4) Joseph R. Lakowicz. *Principles of Fluorescence Spectroscopy*; Lakowicz, J. R., Ed.; Springer US: Boston, MA, 2006. https://doi.org/10.1007/978-0-387-46312-4.

(5) Ma, L.; Yang, F.; Zheng, J. Application of Fluorescence Resonance Energy Transfer in Protein Studies. *J. Mol. Struct.* **2014**, *1077*, 87–100. https://doi.org/https://doi.org/10.1016/j.molstruc.2013.12.071.

(6) Szöllosi, J.; Nagy, P.; Sebestyén, Z.; Damjanovich, S.; Park, J. W.; Mátyus, L. Applications of Fluorescence Resonance Energy Transfer for Mapping Biological

Membranes. *Rev. Mol. Biotechnol.* **2002**, *82* (3), 251–266. https://doi.org/https://doi.org/10.1016/S1389-0352(01)00041-1.

(7) Hong, S.; Samson, A. A. S.; Song, J. M. Application of Fluorescence Resonance Energy Transfer to Bioprinting. *TrAC Trends Anal. Chem.* **2020**, *122*, 115749. https://doi.org/https://doi.org/10.1016/j.trac.2019.115749.

(8) Wu, L.; Huang, C.; Emery, B. P.; Sedgwick, A. C.; Bull, S. D.; He, X. P.; Tian, H.; Yoon, J.; Sessler, J. L.; James, T. D. Förster Resonance Energy Transfer (FRET)-Based Small-Molecule Sensors and Imaging Agents. *Chem. Soc. Rev.* **2020**, *49* (15), 5110–5139. https://doi.org/10.1039/C9CS00318E.

(9) Bradac, C.; Xu, Z. Q.; Aharonovich, I. Quantum Energy and Charge Transfer at Two-Dimensional Interfaces. *Nano Lett.* **2021**, *21* (3), 1193–1204. https://doi.org/10.1021/acs.nanolett.0c04152.

(10) Gupta, S.; Shirodkar, S. N.; Kutana, A.; Yakobson, B. I. In Pursuit of 2D Materials for Maximum Optical Response. *ACS Nano* **2018**, *12* (11), 10880–10889. https://doi.org/10.1021/acsnano.8b03754.

(11) Zhang, C.; Chuu, C. P.; Ren, X.; Li, M. Y.; Li, L. J.; Jin, C.; Chou, M. Y.; Shih, C. K. Interlayer Couplings, Moiré Patterns, and 2D Electronic Superlattices in MoS2/WSe2 Hetero-Bilayers. *Sci. Adv.* **2017**, *3* (1). https://doi.org/10.1126/sciadv.1601459.

(12) Rivera, P.; Yu, H.; Seyler, K. L.; Wilson, N. P.; Yao, W.; Xu, X. Interlayer Valley Excitons in Heterobilayers of Transition Metal Dichalcogenides. *Nat. Nanotechnol.* **2018**, *13* (11), 1004–1015. https://doi.org/10.1038/s41565-018-0193-0.

(13) Miller, B.; Steinhoff, A.; Pano, B.; Klein, J.; Jahnke, F.; Holleitner, A.; Wurstbauer, U. Long-Lived Direct and Indirect Interlayer Excitons in van Der Waals Heterostructures. *Nano Lett.* **2017**, *17* (9), 5229–5237. https://doi.org/10.1021/acs.nanolett.7b01304.

(14) Han, W. Perspectives for Spintronics in 2D Materials. *APL Mater.* **2016**, *4* (3), 32401. https://doi.org/10.1063/1.4941712.

(15) Özcelik, V. O.; Azadani, J. G.; Yang, C.; Koester, S. J.; Low, T. Band Alignment of Two-Dimensional Semiconductors for Designing Heterostructures with Momentum Space Matching. *Phys. Rev. B* **2016**, *94* (3), 35125. https://doi.org/10.1103/PhysRevB.94.035125.

(16) Wang, K.; Huang, B.; Tian, M.; Ceballos, F.; Lin, M. W.; Mahjouri-Samani, M.; Boulesbaa, A.; Puretzky, A. A.; Rouleau, C. M.; Yoon, M.; Zhao, H.; Xiao, K.; Duscher, G.; Geohegan, D. B. Interlayer Coupling in Twisted WSe2/WS2 Bilayer Heterostructures Revealed by Optical Spectroscopy. *ACS Nano* **2016**, *10* (7), 6612–6622. https://doi.org/10.1021/acsnano.6b01486.

(17) Ceballos, F.; Bellus, M. Z.; Chiu, H. Y.; Zhao, H. Ultrafast Charge Separation and Indirect Exciton Formation in a MoS2–MoSe2 van Der Waals Heterostructure. *ACS Nano* **2014**, *8* (12), 12717–12724. https://doi.org/10.1021/nn505736z.

(18) Peng, B.; Yu, G.; Liu, X.; Liu, B.; Liang, X.; Bi, L.; Deng, L.; Sum, T. C.; Loh, K. P. Ultrafast Charge Transfer in MoS 2 /WSe 2 p–n Heterojunction. *2D Mater.* **2016**, *3* (2), 25020. https://doi.org/10.1088/2053-1583/3/2/025020.

(19) Kozawa, D.; Carvalho, A.; Verzhbitskiy, I.; Giustiniano, F.; Miyauchi, Y.; Mouri, S.; Castro Neto, A. H.; Matsuda, K.; Eda, G. Evidence for Fast Interlayer Energy Transfer in MoSe2/WS2 Heterostructures. *Nano Lett.* **2016**, *16* (7), 4087–4093. https://doi.org/10.1021/acs.nanolett.6b00801.

(20) Saha, D.; Varghese, A.; Lodha, S. Atomistic Modeling of van Der Waals Heterostructures with Group-6 and Group-7 Monolayer Transition Metal Dichalcogenides for Near Infrared/Short-Wave Infrared Photodetection. *ACS Appl. Nano Mater.* **2020**, *3* (1), 820–829. https://doi.org/10.1021/acsanm.9b02342.

(21) Lin, T. N.; Huang, L. T.; Shu, G. W.; Yuan, C. T.; Shen, J. L.; Lin, C. A. J.; Chang, W. H.; Chiu, C. H.; Lin, D. W.; Lin, C. C.; Kuo, H. C. Distance Dependence of Energy Transfer from InGaN Quantum Wells to Graphene Oxide. *Opt. Lett.* **2013**, *38* (15),

2897–2899. https://doi.org/10.1364/OL.38.002897.

(22) Itskos, G.; Heliotis, G.; Lagoudakis, P. G.; Lupton, J.; Barradas, N. P.; Alves, E.; Pereira, S.; Watson, I. M.; Dawson, M. D.; Feldmann, J.; Murray, R.; Bradley, D. D. C. Efficient Dipole-Dipole Coupling of Mott-Wannier and Frenkel Excitons in (Ga,In)N Quantum Well/Polyfluorene Semiconductor Heterostructures. *Phys. Rev. B* **2007**, *76* (3), 35344. https://doi.org/10.1103/PhysRevB.76.035344.

(23) Taghipour, N.; Hernandez Martinez, P. L.; Ozden, A.; Olutas, M.; Dede, D.; Gungor, K.; Erdem, O.; Perkgoz, N. K.; Demir, H. V. Near-Unity Efficiency Energy Transfer from Colloidal Semiconductor Quantum Wells of CdSe/CdS Nanoplatelets to a Monolayer of MoS2. *ACS Nano* **2018**, *12* (8), 8547–8554. https://doi.org/10.1021/acsnano.8b04119.

(24) Liu, E.; Fu, Y.; Wang, Y.; Feng, Y.; Liu, H.; Wan, X.; Zhou, W.; Wang, B.; Shao, L.; Ho, C. H.; Huang, Y. S.; Cao, Z.; Wang, L.; Li, A.; Zeng, J.; Song, F.; Wang, X.; Shi, Y.; Yuan, H.; Hwang, H. Y.; Cui, Y.; Miao, F.; Xing, D. Integrated Digital Inverters Based on Two-Dimensional Anisotropic ReS2 Field-Effect Transistors. *Nat. Commun.* **2015**, *6* (1), 6991. https://doi.org/10.1038/ncomms7991.

(25) Wolverson, D.; Crampin, S.; Kazemi, A. S.; Ilie, A.; Bending, S. J. Raman Spectra of Monolayer, Few-Layer, and Bulk ReSe2: An Anisotropic Layered Semiconductor. *ACS Nano* **2014**, *8* (11), 11154–11164. https://doi.org/10.1021/nn5053926.

(26) Tongay, S.; Suh, J.; Ataca, C.; Fan, W.; Luce, A.; Kang, J. S.; Liu, J.; Ko, C.; Raghunathanan, R.; Zhou, J.; Ogletree, F.; Li, J.; Grossman, J. C.; Wu, J. Defects Activated Photoluminescence in Two-Dimensional Semiconductors: Interplay between Bound, Charged and Free Excitons. *Sci. Rep.* **2013**, *3* (1), 2657. https://doi.org/10.1038/srep02657.

(27) Tonndorf, P.; Schmidt, R.; Böttger, P.; Zhang, X.; Börner, J.; Liebig, A.; Albrecht, M.; Kloc, C.; Gordan, O.; Zahn, D. R. T.; Michaelis de Vasconcellos, S.; Bratschitsch, R.

Photoluminescence Emission and Raman Response of Monolayer MoS2, MoSe2, and WSe2. *Opt. Express* **2013**, *21* (4), 4908–4916. https://doi.org/10.1364/OE.21.004908.

(28) Mohamed, N. B.; Lim, H. E.; Wang, F.; Koirala, S.; Mouri, S.; Shinokita, K.; Miyauchi, Y.; Matsuda, K. Long Radiative Lifetimes of Excitons in Monolayer Transition-Metal DichalcogenidesMX2(M= Mo, W;X= S, Se). *Appl. Phys. Express* **2017**, *11* (1), 15201. https://doi.org/10.7567/apex.11.015201.

(29) He, R.; Yan, J. A.; Yin, Z.; Ye, Z.; Ye, G.; Cheng, J.; Li, J.; Lui, C. H. Coupling and Stacking Order of ReS2 Atomic Layers Revealed by Ultralow-Frequency Raman Spectroscopy. *Nano Lett.* **2016**, *16* (2), 1404–1409. https://doi.org/10.1021/acs.nanolett.5b04925.

(30) Chenet, D. A.; Aslan, O. B.; Huang, P. Y.; Fan, C.; van der Zande, A. M.; Heinz, T. F.; Hone, J. C. In-Plane Anisotropy in Mono- and Few-Layer ReS2 Probed by Raman Spectroscopy and Scanning Transmission Electron Microscopy. *Nano Lett.* **2015**, *15* (9), 5667–5672. https://doi.org/10.1021/acs.nanolett.5b00910.

(31) Yang, A.; Blancon, J. C.; Jiang, W.; Zhang, H.; Wong, J.; Yan, E.; Lin, Y. R.; Crochet, J.; Kanatzidis, M. G.; Jariwala, D.; Low, T.; Mohite, A. D.; Atwater, H. A. Giant Enhancement of Photoluminescence Emission in WS2-Two-Dimensional Perovskite Heterostructures. *Nano Lett.* **2019**, *19* (8), 4852–4860. https://doi.org/10.1021/acs.nanolett.8b05105.

(32) He, Z.; Xu, W.; Zhou, Y.; Wang, X.; Sheng, Y.; Rong, Y.; Guo, S.; Zhang, J.; Smith, J. M.; Warner, J. H. Biexciton Formation in Bilayer Tungsten Disulfide. *ACS Nano* **2016**, *10* (2), 2176–2183. https://doi.org/10.1021/acsnano.5b06678.

(33) Aslan, O. B.; Chenet, D. A.; van der Zande, A. M.; Hone, J. C.; Heinz, T. F. Linearly Polarized Excitons in Single- and Few-Layer ReS2 Crystals. *ACS Photonics* **2016**, *3* (1), 96–101. https://doi.org/10.1021/acsphotonics.5b00486.

(34) Raja, A.; Chaves, A.; Yu, J.; Arefe, G.; Hill, H. M.; Rigosi, A. F.; Berkelbach, T. C.;

Nagler, P.; Schüller, C.; Korn, T.; Nuckolls, C.; Hone, J.; Brus, L. E.; Heinz, T. F.; Reichman, D. R.; Chernikov, A. Coulomb Engineering of the Bandgap and Excitons in Two-Dimensional Materials. *Nat. Commun.* **2017**, *8* (1), 15251. https://doi.org/10.1038/ncomms15251.

(35) Rigosi, A. F.; Hill, H. M.; Li, Y.; Chernikov, A.; Heinz, T. F. Probing Interlayer Interactions in Transition Metal Dichalcogenide Heterostructures by Optical Spectroscopy: MoS2/WS2 and MoSe2/WSe2. *Nano Lett.* **2015**, *15* (8), 5033–5038. https://doi.org/10.1021/acs.nanolett.5b01055.

(36) Le, C. T.; Clark, D. J.; Ullah, F.; Senthilkumar, V.; Jang, J. I.; Sim, Y.; Seong, M. J.; Chung, K.-H.; Park, H.; Kim, Y. S. Nonlinear Optical Characteristics of Monolayer MoSe2. *Ann. Phys.* **2016**, *528* (7–8), 551–559. https://doi.org/10.1002/andp.201600006.

(37) Tongay, S.; Sahin, H.; Ko, C.; Luce, A.; Fan, W.; Liu, K.; Zhou, J.; Huang, Y. S.; Ho, C. H.; Yan, J.; Ogletree, D. F.; Aloni, S.; Ji, J.; Li, S.; Li, J.; Peeters, F. M.; Wu, J. Monolayer Behaviour in Bulk ReS2 Due to Electronic and Vibrational Decoupling. *Nat. Commun.* **2014**, *5* (1), 3252. https://doi.org/10.1038/ncomms4252.

(38) Mohamed, N. B.; Shinokita, K.; Wang, X.; Lim, H. E.; Tan, D.; Miyauchi, Y.; Matsuda, K. Photoluminescence Quantum Yields for Atomically Thin-Layered ReS2: Identification of Indirect-Bandgap Semiconductors. *Appl. Phys. Lett.* **2018**, *113* (12), 121112. https://doi.org/10.1063/1.5037116.

(39) Tomita, A.; Shah, J.; Knox, R. S. Efficient Exciton Energy Transfer between Widely Separated Quantum Wells at Low Temperatures. *Phys. Rev. B* **1996**, *53* (16), 10793–10803. https://doi.org/10.1103/PhysRevB.53.10793.

(40) Dandu, M.; Biswas, R.; Das, S.; Kallatt, S.; Chatterjee, S.; Mahajan, M.; Raghunathan, V.; Majumdar, K. Strong Single- and Two-Photon Luminescence Enhancement by Nonradiative Energy Transfer across Layered Heterostructure. *ACS Nano* **2019**, *13* (4),

4795–4803. https://doi.org/10.1021/acsnano.9b01553.

(41) Lin, Y.; Ling, X.; Yu, L.; Huang, S.; Hsu, A. L.; Lee, Y. H.; Kong, J.; Dresselhaus, M. S.; Palacios, T. Dielectric Screening of Excitons and Trions in Single-Layer MoS2. *Nano Lett.* **2014**, *14* (10), 5569–5576. https://doi.org/10.1021/nl501988y.

(42) Gehlmann, M.; Aguilera, I.; Bihlmayer, G.; Nemšák, S.; Nagler, P.; Gospodarič, P.; Zamborlini, G.; Eschbach, M.; Feyer, V.; Kronast, F.; Młyńczak, E.; Korn, T.; Plucinski, L.; Schüller, C.; Blügel, S.; Schneider, C. M. Direct Observation of the Band Gap Transition in Atomically Thin ReS2. *Nano Lett.* **2017**, *17* (9), 5187–5192. https://doi.org/10.1021/acs.nanolett.7b00627.

(43) Ugeda, M. M.; Bradley, A. J.; Shi, S. F.; da Jornada, F. H.; Zhang, Y.; Qiu, D. Y.; Ruan, W.; Mo, S. K.; Hussain, Z.; Shen, Z. X.; Wang, F.; Louie, S. G.; Crommie, M. F. Giant Bandgap Renormalization and Excitonic Effects in a Monolayer Transition Metal Dichalcogenide Semiconductor. *Nat. Mater.* **2014**, *13* (12), 1091–1095. https://doi.org/10.1038/nmat4061.

(44) Yuan, J.; Najmaei, S.; Zhang, Z.; Zhang, J.; Lei, S.; Ajayan, P. M.; Yakobson, B. I.; Lou, J. Photoluminescence Quenching and Charge Transfer in Artificial Heterostacks of Monolayer Transition Metal Dichalcogenides and Few-Layer Black Phosphorus. *ACS Nano* **2015**, *9* (1), 555–563. https://doi.org/10.1021/nn505809d.

(45) Bellus, M. Z.; Li, M.; Lane, S. D.; Ceballos, F.; Cui, Q.; Zeng, X. C.; Zhao, H. Type-I van Der Waals Heterostructure Formed by MoS2 and ReS2 Monolayers. *Nanoscale Horiz.* **2017**, *2* (1), 31–36. https://doi.org/10.1039/C6NH00144K.

(46) Merkl, P.; Mooshammer, F.; Steinleitner, P.; Girnghuber, A.; Lin, K. Q.; Nagler, P.; Holler, J.; Schüller, C.; Lupton, J. M.; Korn, T.; Ovesen, S.; Brem, S.; Malic, E.; Huber, R. Ultrafast Transition between Exciton Phases in van Der Waals Heterostructures. *Nat. Mater.* **2019**, *18* (7), 691–696. https://doi.org/10.1038/s41563-019-0337-0.

(47) Hong, X.; Kim, J.; Shi, S. F.; Zhang, Y.; Jin, C.; Sun, Y.; Tongay, S.; Wu, J.; Zhang, Y.; Wang, F. Ultrafast Charge Transfer in Atomically Thin MoS2/WS2 Heterostructures. *Nat. Nanotechnol.* **2014**, *9* (9), 682–686. https://doi.org/10.1038/nnano.2014.167.

(48) Park, Y.; Han, S. W.; Chan, C. C. S.; Reid, B. P. L.; Taylor, R. A.; Kim, N.; Jo, Y.; Im, H.; Kim, K. S. Interplay between Many Body Effects and Coulomb Screening in the Optical Bandgap of Atomically Thin MoS2. *Nanoscale* **2017**, *9* (30), 10647–10652. https://doi.org/10.1039/C7NR01834G.

(49) Barnes, W. L. Fluorescence near Interfaces: The Role of Photonic Mode Density. *J. Mod. Opt.* **1998**, *45* (4), 661–699. https://doi.org/10.1080/09500349808230614.

(50) Li, M.; Chen, J. S.; Cotlet, M. Efficient Light Harvesting Biotic–Abiotic Nanohybrid System Incorporating Atomically Thin van Der Waals Transition Metal Dichalcogenides. *ACS Photonics* **2019**, *6* (6), 1451–1457. https://doi.org/10.1021/acsphotonics.9b00090.

(51) Tanoh, A. O. A.; Gauriot, N.; Delport, G.; Xiao, J.; Pandya, R.; Sung, J.; Allardice, J.; Li, Z.; Williams, C. A.; Baldwin, A.; Stranks, S. D.; Rao, A. Directed Energy Transfer from Monolayer WS2 to Near-Infrared Emitting PbS–CdS Quantum Dots. *ACS Nano* **2020**, *14* (11), 15374–15384. https://doi.org/10.1021/acsnano.0c05818.

(52) Lyo, S. K. Energy Transfer of Excitons between Quantum Wells Separated by a Wide Barrier. *Phys. Rev. B* **2000**, *62* (20), 13641–13656. https://doi.org/10.1103/PhysRevB.62.13641.

(53) Stavola, M.; Dexter, D. L.; Knox, R. S. Electron-Hole Pair Excitation in Semiconductors via Energy Transfer from an External Sensitizer. *Phys. Rev. B* **1985**, *31* (4), 2277–2289. https://doi.org/10.1103/PhysRevB.31.2277.

(54) Wu, L.; Chen, Y.; Zhou, H.; Zhu, H. Ultrafast Energy Transfer of Both Bright and Dark Excitons in 2D van Der Waals Heterostructures Beyond Dipolar Coupling. *ACS*

*Nano* **2019**, *13* (2), 2341–2348. https://doi.org/10.1021/acsnano.8b09059.

(55) He, K.; Kumar, N.; Zhao, L.; Wang, Z.; Mak, K. F.; Zhao, H.; Shan, J. Tightly Bound Excitons in Monolayer WSe2. *Phys. Rev. Lett.* **2014**, *113* (2), 26803. https://doi.org/10.1103/PhysRevLett.113.026803.

(56) Kowerko, D.; Krause, S.; Amecke, N.; Abdel-Mottaleb, M.; Schuster, J.; Von Borczyskowski, C. Identification of Different Donor-Acceptor Structures via Förster Resonance Energy Transfer (FRET) in Quantum-Dot-Perylene Bisimide Assemblies. *Int. J. Mol. Sci.* **2009**, *10* (12), 5239–5256. https://doi.org/10.3390/ijms10125239.

(57) Gu, J.; Liu, X.; Lin, E.; Lee, Y. H.; Forrest, S. R.; Menon, V. M. Dipole-Aligned Energy Transfer between Excitons in Two-Dimensional Transition Metal Dichalcogenide and Organic Semiconductor. *ACS Photonics* **2018**, *5* (1), 100–104. https://doi.org/10.1021/acsphotonics.7b00730.